\documentstyle[11pt]{article}

\begin{document}
\begin{titlepage}
\begin{center}
{\Large Theoretical Physics Institute}\\
{\Large University of Minnesota}
\end{center}
\vskip0.5cm
\begin{flushright}
TPI--MINN--93/46--T\\
CBPF--NF--058/93\\
DFTT 58/93\\
October 1993
\end{flushright}
\vskip0.5cm
\begin{center}
{\Large\bf Capillary Wave Approach to Order-Order Fluid}
\vskip0.2cm
{\Large\bf Interfaces in the 3D Three-State Potts Model}
\end{center}
\begin{center}
{\large Paolo Provero}\\
{\it Theoretical Physics Institute}\\
{\it University of Minnesota}\\
{\it Minneapolis, MN 55455, USA}
\footnote{e--mail: paolo@physics.spa.umn.edu}
\end{center}
\vskip0.2cm
\begin{center}
{\large Stefano Vinti}\\
{\it CBPF}\\
{\it Centro Brasileiro de Pesquisas Fisicas}\\
{\it Rua Dr. Xavier Sigaud 150}\\
{\it 22290 Rio de Janeiro, Brasil}\\
{\it and}\\
{\it Dipartimento di Fisica Teorica}\\
{\it Universit\`a di Torino}\\
{\it Via P. Giuria 1}\\
\it{10125 Torino, Italia}
\footnote{e--mail: vinti@to.infn.it}
\end{center}
\vskip0.5cm

\begin{abstract}
The physics of fluid interfaces between domains of different
magnetization in the ordered phase of the 3D three--state Potts model
is studied by means of a Monte Carlo simulation.
It is shown that finite--size effects in the interface free energy
are well described by the capillary wave model at two loop order,
supporting the idea of the universality of this description of
fluid interfaces in 3D statistical models.
\end{abstract}
\end{titlepage}

\section {Introduction}

It is  well known that 3D spin systems on finite volumes show domain
walls separating coexisting phases which behave as {\em fluid interfaces}
between the critical and the roughening temperature.
The finite-size effects in the free energy of a fluid interface
are dominated by long-wavelength fluctuations and a correct physical
description of the critical properties of the surface cannot neglect
their contributions \cite{GeF}.

While below the roughening temperature, where the interfaces are almost
rigid, a microscopical approach can be taken (see \cite{Borgs} and
references therein), above it one is forced to assume an effective model
describing the collective degrees of freedom of rough interfaces.

The {\em capillary wave model} (CWM) \cite{BLS} in its simplest
formulation, which we will follow, assumes an effective hamiltonian
proportional to the area of the surface.

It has been recently shown \cite{cgv} that rather strong finite-size
effects, depending on the shape of the lattice, are well described in
terms of the {\em one--loop}
or gaussian approximation to the CWM: its predictions
have been tested with high accuracy in the scaling region of the 3D
Ising model.
These one--loop corrections
depend only on one adimensional parameter, namely on the asymmetry
$z=R/T$ of the transverse sizes of an elongated lattice $R\times T\times
L$ ($L\gg R,T$) with periodic boundary conditions taken in all
directions \cite{IZ,bunk,cgv}.

As it has been already pointed out \cite{cgv}, higher order corrections
to the gaussian model can be taken into account
to verify the CWM beyond the one--loop approximation.
These higher order corrections can provide a more stringent test
of the CWM model: in fact, while many different effective hamiltonians
reduce to the gaussian form at one--loop level \cite{luscher},
they differ in the form of two-- and higher--loop corrections.
In this paper we apply the CWM in the two--loop approximation to the
finite--size behavior of order--order rough interfaces in the 3D
three-state Potts model.
\par
The two--loop contributions do not
depend only on the asymmetry parameter $z$ but also on an adimensional
parameter proportional to the minimal area of the surface, namely
$\sigma R T$, $\sigma$ being the (reduced) interface tension.
An important consequence of this fact is that the well known finite-size
behavior \cite{FeP} of the energy splitting $E$ occurring between
vacua on finite volumes, has no longer, at two--loop, the {\em classical}
functional form
\begin{equation}
 E_{cl}(R) \propto e^{-\sigma R^2}
\end{equation}
not even for symmetric ($R=T$) lattices. This must be contrasted with
what happens at one--loop level, where, due to the scale--invariance of
the one--loop contributions,
one has to consider asymmetric
lattices to find deviations from the classical functional form
\cite{cgv}.
\par
To verify the CWM we study the finite-size behavior of
rough order-order interfaces of the 3D three-state Potts model by
means of Monte Carlo numerical simulations.

This model is of great interest because of its well known
connection with 4D $SU(3)$ pure Yang--Mills theory at finite
temperature \cite{SeY}.
In fact one can assume an effective action of a 3D
spin model with short--range, ferromagnetic interaction
to describe the finite temperature
deconfinement transition of QCD in the limit of infinite quark
masses \cite{FUO}.
For this reason the Potts model has been extensively investigated and,
in particular, the properties of order-order and order-disorder
interfaces have been already studied \cite{KeP,WdT,Gretall}, one of the
main goals being the evaluation of the corresponding interface
tensions.

In this paper we show that the knowledge of the functional form of the
fluid interface free energy on finite volumes,
including the capillary wave contributions,
enables one to estimate the order-order interface tension with high
precision.
Moreover, the good agreement we find with the theoretical predictions
of the CWM model at two--loop order strongly supports the idea
of the universality of this description of rough interfaces in 3D
statistical models.
\par
The paper is structured as follows: in Sec. 2. we describe the CWM and
its one--loop and two--loop approximations. In Sec. 3. we compare the
results of MC simulations with the CWM predictions. Sec. 4 is devoted
to some concluding remarks.
\section{The capillary wave model}

According to the CWM, the interface between
two domains of different magnetization in the ordered phase of a
3D spin system, above the roughening temperature,
is described by the partition function

\begin{equation}
Z_{cw}=\int \left[ Dx\right]\exp\left\{-\sigma A\left[ x\right]
\right\}~~,
\label{zeta}
\end{equation}

\noindent
where the single-valued function $x(r,t)$ describes the displacement
from the equilibrium position of the interface, $\sigma$ is the reduced
(order-order) interface tension and $A[x]$ is the area of
the interface

\begin{equation}
A\left[ x\right]=\int_0^{R} dr
\int_0^{T}dt~ \sqrt{1+\left(\frac{\partial x}{\partial r}\right)^2
+\left(\frac{\partial x}{\partial t}\right)^2}~~.
\label{acw}
\end{equation}

\noindent
It should be mentioned that (\ref{acw}) coincides with the
Nambu-Goto string action in $D=3$ in a particular gauge.
Eq. (\ref{acw}) is not expected to be the
{\em exact} action describing fluid interfaces but at least
the dominant contribution\footnote
{The problems arising in the quantization of
non--critical strings are known to disappear asymptotically at
large distances \cite{olesen}.}: as we will show this is indeed the case.

\par
To compare the predictions of the CWM with numerical
results from Monte Carlo (MC) simulations, we have chosen 3D lattices
of $R\times T\times L$ sites, with $L\gg R,T$, and
periodic boundary conditions in each direction.
This particular choice of the lattice shape allows one to consider
only interfaces orthogonal to the elongated ($L$) direction,
the probability of having interfaces orthogonal to the other directions
being negligible.
The 2D field $x(r,t)$ is therefore defined on the rectangle $(r,t) \in
\left[0,R\right]\times \left[0,T\right]$ with opposite edges
identified, i.e. on a torus.

The partition function (\ref{zeta}) cannot be computed exactly,
but it is possible to express it as an expansion in powers of
the adimensional parameter $(\sigma RT)^{-1}$: the two--loop expansion
of $Z_{cw}$ can then be written as
\begin{equation}
Z_{cw}(R,T) \propto e^{-\sigma RT}~
Z_q^{(1~  loop)}\left( \frac{R}{T}\right)\cdot
Z_q^{(2~  loop)}\left(R,T\right)~~.
\label{gen}
\end{equation}
The one--loop contribution (namely the gaussian approximation), obtained
retaining only the quadratic term in the expansion of (\ref{acw}),
is nothing else than the exact partition function of a 2D conformal
invariant free boson on a torus of modular parameter
$\tau=i\frac{R}{T}$ \cite{IZ,bunk,cgv}

\begin{equation}
Z_q^{(1~  loop)}\left( \frac{R}{T}\right)=
\sqrt{\frac{T}{R}}\left|\eta\left(i\frac{R}{T}\right)\right|^{-2}~~,
\label{1loop}
\end{equation}
while the two--loop term can be calculated perturbatively expanding
(\ref{acw}) at the next--to--leading order \cite{DF}
\begin{eqnarray}
Z_q^{(2~loop)}(R,T)&=&
1+\frac{1}{2\sigma RT}\left\{
\left[\frac{\pi}{6} \frac{R}{T} E_2\left(i\frac{R}{T}\right)\right]^2 -
\frac{\pi}{6} \frac{R}{T} E_2\left(i\frac{R}{T}\right) +
\frac{3}{4}\right\}\nonumber\\
&&  +~~~~ O\left[\left(\sigma RT\right)^{-2}\right]~~.
\label{2loop}
\end{eqnarray}

\noindent
The two functions $\eta$ and $E_2$ appearing above are respectively
the Dedekind function and the second Eisenstein series:

\begin{eqnarray}
\eta(\tau)&=&q^{1/24}\prod_{n=1}^{\infty}\left(1-q^n\right),\quad\quad
q\equiv \exp(2\pi i \tau)\nonumber\\
E_2(\tau)&=&1-24\sum_{n=1}^{\infty}\frac{n~ q^n}{1-q^n}~~~.
\nonumber
\end{eqnarray}
\par
The three--state Potts model is defined by the partition function
\begin{equation}
Z=\sum_{\left\{\sigma_i\right\}}\exp \left\{-\beta\sum_{i,\hat{\mu}}
\left[1-Re(\sigma_i^*\sigma_{i+\hat{\mu}})\right]\right\}
\end{equation}
where the variables $\sigma_i$ are defined on a three--dimensional
hypercubic lattice and take the values
\begin{equation}
\sigma_i=\exp\left(\frac{2\pi i n_i}{3}\right) \qquad n_i=0,1,2~~.
\end{equation}
In the termodynamic limit the Potts model is known to undergo a (weak)
 first--order phase transition
at $\beta_c=0.36708(2)$ \cite{GKP} and the roughening temperature
can be estimate to be $\beta_r \sim 0.93$ \cite{DeZ}.
For $\beta> \beta_c$ the $Z_3$ symmetry is spontaneously broken and the
three ordered phases coexist, while at $\beta=\beta_c$ also the disordered
phase coexists with the previous ones.
\par
In the finite {\em cylindric} geometry we are considering
spontaneous symmetry breaking at low temperature cannot occur:
the degeneracy of the ground state is removed,
the energy of the symmetric, $Z_3$ invariant, ground state being
separated by an energy splitting $E$ from the two degenerate
mixed-symmetry states.
\par
The energy splitting is due to tunneling between the phases and
is directly linked to the free energy of the interface \cite{FeP}.
According to the CWM, for $\beta_r>\beta>\beta_c$, we assume
\cite{GeF,BLS,cgv}, $R\ge T$,

\begin{eqnarray}
E(R,T) &=& C~ e^{-\sigma RT}~
Z_q^{(1~loop)}\left( \frac{R}{T}\right)\cdot
Z_q^{(2~loop)}\left(R,T\right)
\label{e1}\\
C &=& \frac{\delta }{Z_q^{(1~loop)}\left( 1\right)}\nonumber
\end{eqnarray}
where $\delta$ is an unpredicted constant and a convenient
normalization has been chosen.

\par
We would like to stress that the two--loop contribution (\ref{2loop})
does not depend only on the ratio $z=R/T$, like the one--loop term
(\ref{1loop}), but also on the minimal area $A_m=RT$.
If we put $z=1$, $A_m=R^2$, in (\ref{e1}) we obtain

\begin{equation}
E(R,R)= \delta~ e^{-\sigma R^2}
\left\{1+\frac{1}{2\sigma R^2}
\left[\left(\frac{\pi}{6}f \right)^2-\frac{\pi}{6}f
+\frac{3}{4}\right]\right\}~~,
\label{e2}
\end{equation}
where $f\equiv E_2(i)$.
The {\em classical} formula \cite{FeP}

\begin{equation}
 E_{cl}(R)=\delta ~e^{-\sigma R^2}
\label{e3}
\end{equation}
\noindent
can be recovered only neglecting the two--loop contribution.

The comparison between  formula (\ref{e1}) and
the values of $E$ extracted from MC simulations provides a simple
and stringent way to verify the CWM predictions.

It should be noted that no new free parameters are introduced
within this approach: the formulae (\ref{e1}) and (\ref{e3}) contain the
same number of undetermined parameters, namely $\sigma$ and $\delta$.

\section{Monte Carlo results}

To extract the energy splitting $E$ from MC--generated ensembles
we follow the procedure of \cite{KeM}.
\noindent
Defining the time--slice magnetization
\begin{equation}
S_k\equiv \frac{1}{RT}\sum_{x_1=1}^R\sum_{x_2=1}^T\sigma(x_1,x_2,k)~~,
\end{equation}
\noindent
we compute the correlation function
\begin{equation}
G(k)\equiv\langle S_0 S_k^*\rangle ~~
\end{equation}
where $k=0,1,\dots,\frac{L}{2}~$,
and we extract the transfer matrix low energy levels from the
asymptotic $k$--dependence of $G(k)$
\begin{eqnarray}
G(k) \cdot Z &=&c_0\left\{\exp(-kE)+\exp[-(L-k)E]\right\}\nonumber\\
&&+c_1\left\{\exp(-kE')+\exp[-(L-k)E']\right\}~+~\dots~~\label{fit}\\
Z &=& 1 + 2 e^{-L E} + \dots\label{gk}
\end{eqnarray}
$Z\equiv$ tr $e^{-LH}$ being the partition function (the
next--to--leading energy level $E'$ turns out to be non--negligible
in our range of parameters).
\par
Having so extracted the energy splitting $E$ from the MC data for
different values of the lattice sizes, we can compute
the order-order interface tension $\sigma$ and the constant
$\delta$ by fitting our data with the formula (\ref{e1}).

\par
We have performed our simulations at $\beta=0.3680$,
the longest lattice size being fixed at $L=120$, and the other
sizes varying in the range $9\le T\le 11$, $10\le R\le 36$, ($R\ge T$).
This value of $\beta$ is enough inside the
ordered phase to make highly suppressed the probability
of formation of order-disorder interfaces \cite{KeP}
but presents a correlation length large enough to make the lattice
artifacts negligible and to consider domain walls as fluid interfaces.
\par
The fact that the disordered phase is substantially absent at this $\beta$
can be seen from the histograms of the real part of the magnetization
\cite{GIPGK}
\begin{equation}
Re M\equiv Re \left(\frac{1}{L}\sum_{k=1}^L S_k\right)\nonumber
\end{equation}
as is shown, for example, in Fig.1. The modulus of the magnetization at
this $\beta$ is about $0.44-0.50$ for the
lattice sizes we are considering:
in this figure the projection on the real axis of the ordered phases
are clearly visible while the peak centered at $Re M=0$, which would
signal the presence of the disordered phase, is absent.
\par
Fig. 2 represents a typical distribution of the magnetization
$M$ for a sample of our MC--generated configurations. Most
configurations consist of a single phase or of two phases separated
by two interfaces (the minimum number compatible with periodic
boundary conditions). The single--phase configurations are represented
by the three clusters of points corresponding to the three degenerate vacua;
the two--interface configurations form the straight lines joining these
clusters. Three--interface configurations, which tend to fill uniformly
the interior of the triangle, are clearly visible in Fig. 3:
it corresponds to a $T=10$, $R=20$ lattice and to a larger
probability of having tunneling events (i.e. interfaces), while these
are much more suppressed for the lattice of Fig. 2 ($T=R=18$).

We have used a Swendsen-Wang cluster algorithm \cite{SeW} to perform
our MC simulations. To keep under control correlations in MC
time and cross-correlations between the $G(k)$ observables, we have
systematically scattered our measurements avoiding the measurement of two
different observables at the same MC time. We have made between
$0.6\cdot 10^6$ and $1.8\cdot 10^6$ sweeps for each experiment, depending
on the lattice size, obtaining about $10^3$ data per observable.
However, the covariance matrix turns out to be different from
the diagonal form, which one expects from a sample of statistical
independent data. We have taken this fact into account by including the
covariance matrix in the fitting procedure to formula (\ref{fit})
to extract the energy gap $E$: the results are reported in Tab. 1.
The error on $E$ has been estimated with an ordinary jackknife
procedure.

\vskip0.3cm
\begin{center}
\begin{tabular}{|c|c|c|c|c|c|} \hline
$T$&$R$&$E$ (MC)&$\chi^2$&C.L.&$E$ (CWM) \\ \hline
10&10&0.06399(51)&0.89&65\%&0.06399 \\ \hline
12&12&0.03924(54)&1.07&36\%&0.03882 \\ \hline
14&14&0.02201(54)&0.75&80\%&0.02227 \\ \hline
16&16&0.01234(32)&0.95&52\%&0.01196 \\ \hline
18&18&0.00508(71)&1.03&42\%&0.00598 \\ \hline
20&20&0.00305(68)&0.95&57\%&0.00278 \\ \hline
9&18&0.04108(73)&0.54&97\%&0.04105 \\ \hline
9&21&0.03467(96)&0.99&47\%&0.03528 \\ \hline
9&24&0.0293(11)&1.09&34\%&0.03082 \\ \hline
9&27&0.0283(10)&0.90&61\%&0.02723 \\ \hline
9&30&0.0235(11)&1.16&27\%&0.02427 \\ \hline
9&36&0.0188(13)&0.98&48\%&0.01959 \\ \hline
10&18&0.03127(47)&0.44&99\%&0.03119 \\ \hline
10&20&0.02670(61)&1.01&45\%&0.02710 \\ \hline
10&22&0.02339(34)&0.88&64\%&0.02374 \\ \hline
10&24&0.02049(46)&0.79&75\%&0.02092 \\ \hline
10&26&0.01960(47)&1.02&44\%&0.01854 \\ \hline
10&28&0.01644(70)&0.83&69\%&0.01650 \\ \hline
10&30&0.01473(88)&0.84&80\%&0.01473 \\ \hline
11&20&0.02024(75)&1.17&25\%&0.02049 \\ \hline
11&22&0.01766(78)&1.05&39\%&0.01735 \\ \hline
11&24&0.01495(61)&0.77&77\%&0.01479 \\ \hline
11&26&0.01292(69)&0.67&84\%&0.01267 \\ \hline
11&28&0.01024(91)&0.66&87\%&0.01090 \\ \hline
11&32&0.0078(12)&0.85&73\%&0.00815 \\ \hline
\end{tabular}
\end{center}
\begin{center}
{\bf Tab. 1.} {\it The values of $E$ are reported with the $\chi^2$
per degree of freedom and the confidence levels, as obtained from the
fit of $G(k)$ with formula (\ref{fit}).
The values in the last column are obtained from the best fit of all
data to formula (\ref{e1}). The same data are plotted in Fig. 4.}
\end{center}
\vskip0.3cm
Fitting our results for the energy gaps $E$ with the CWM formula
(\ref{e1}) we obtain the following values of the interface tension and
of the constant:
\begin{eqnarray}
&&\sigma=0.009912(75)\nonumber\\
&&\delta=0.1377(19)\nonumber
\end{eqnarray}
with a $\chi^2$ per degree of freedom and a confidence level
\begin{equation}
\chi^2=0.73\qquad{\rm C. L.}=82\%
\end{equation}
thus confirming the accuracy of the CWM. In Tab. 1 the
MC results for $E$ are compared with the predictions of
formula (\ref{e1}) in which the best--fit values of $\delta$
and $\sigma$ have been substituted. This comparison is represented
graphically in Fig. 4.

The importance of the inclusion of the two--loop contributions can
be seen by fitting the MC data with the classical formula
\begin{equation}
 E(R,T)=\delta ~e^{-\sigma RT}
\label{class}
\end{equation}
and with the one--loop approximation ($R\ge T$)
\begin{equation}
E(R,T)=\delta~e^{-\sigma RT}~\frac{Z_q^{(1\ loop)}\left(\frac{R}{T}
\right)}{Z_q^{1\ loop}(1)}\label{1lupo}~~.
\end{equation}
\noindent
In the former case we obtain $\chi^2/{\rm d.o.f.}=36.3$, in the latter
$\chi^2/{\rm d.o.f}=3.60$: the two--loop correction must be included to obtain
a good agreement with numerical data.
\par
We have already noted that the two--loop corrections
affect the value of $E$ also for symmetric ($T=R$) lattices (cfr. (\ref{e2})),
in contrast to what happens for the scale--invariant one--loop
contribution (\ref{1loop}) \cite{cgv}.
Indeed, the importance of including two--loop corrections can be
seen by fitting only the energy gaps $E$ obtained on symmetric
lattices ($T=R$): using the two--loop
expression (\ref{e2}) we obtain $\chi^2=0.79$, while the classical
formula (\ref{e3}) gives $\chi^2=1.48$.
The results of all these fits are summarized in Tab. 2.

We would like to stress the remarkable stability of the
results obtained with the two--loop approximation fitting all
the data or only the symmetric ones. This can be seen comparing the
values of $\sigma$ and $\delta$ given in the first and fourth line
of Tab. 2.

\vskip0.5cm
\begin{center}
\begin{tabular}{|c|c|c|c|c|c|} \hline
Data&Approx.&$\sigma$&$\delta$&$\chi^2$&C.L.\\ \hline
all&2--loop&0.009912(75)&0.1377(19)&0.73&0.82 \\ \hline
all&1--loop&0.010053(75)&0.1724(23)&3.60&0.00 \\ \hline
all&class.&0.008092(75)&0.1395(19)&36.3&0.00 \\ \hline
$T=R$&2--loop&0.00981(14)&0.1361(26)&0.79&0.53 \\ \hline
$T=R$&class.&0.01075(14)&0.1866(36)&1.48&0.20 \\ \hline
$T=11$&2--loop&0.00997(69)&0.140(24)&0.26&0.90 \\ \hline
$T=11$&1--loop&0.00965(69)&0.151(26)&0.22&0.92 \\ \hline
\end{tabular}
\end{center}
\begin{center}
{\bf Tab. 2.} {\it Results of the fit of $E$ with two--loop,
one--loop and classical approximations of the CWM,
considering all values of $(R,T)$, symmetric $(R,R)$ lattices or
(R,T=11) lattices.}
\end {center}
\vskip0.3cm
On the other hand, the result obtained using the classical formula
(\ref{class}) is  not
compatible with the previous ones even using only the symmetric data,
as it is shown in the fifth line of the same table.
The best fit curve obtained from (\ref{class}) in the latter case is
plotted in Fig. 5 were also the  "asymmetric" MC data are
reported for comparison.
\par
We would also like to observe that a good agreement with the
one--loop approximation of the CWM can be obtained \cite{cgv}
if one considers low values of ratios $z=R/T$ and high values of the
minimal area $A_m=RT$, i.e. where the two--loop contribution (\ref{2loop})
are maximally suppressed. This is show in the last two lines
of Tab. 2 and in Fig.6.
\section{Conclusions}

In this paper we have shown that the CWM in the two--loop
approximation provides an excellent description of order--order
interfaces in the 3D three--state Potts model. This result,
together with the corresponding one for the 3D Ising model
\cite{cgv}, strongly supports the hypothesis of the universality of the
CWM description of interface physics in 3D statistical models.
\par
It is worth stressing again that the CWM corrections (\ref{e1}) to
the finite--size behavior of the interface free energy do not introduce
any new free parameters with respect to the "classical" picture
(\ref{class}).
\par
Besides the intrinsic physical interest of this picture,
it should be noted that the CWM provides an accurate description
of finite--size corrections to the free energy of rough interfaces,
thus enabling one to extract correct informations about physical
observables from finite lattices of different geometries, as it has
been recently shown \cite{IKKRY}.
\vskip0.5cm
{\bf Acknowledgements}
\vskip0.5cm
We are grateful to M. Caselle and F. Gliozzi for many useful
suggestions.
One of us (S.V.) would like to thank S. Alves Dias, M. G. do Amaral,
C. A. Linhares, J. Mignaco and M. A. do Rego Monteiro for
stimulating discussions and the CBPF for the kind and warm
hospitality in Rio de Janeiro.
The work of P.P. is supported by an INFN grant.
The work of S.V. in Rio de Janeiro is supported by an CNPq grant.

\newpage
\begin{center}
{\bf Figure Captions}
\end{center}
\vskip0.5cm
{\bf Fig. 1.} {\it Histogram of the real part of the magnetization for a
typical MC ensemble (in this case $T=20$, $R=20$). The absence of
a peak in $Re\ M=0$ indicates that the disordered phase does not coexist
with the ordered ones at our $\beta=0.3680$.}
\vskip0.5cm
{\bf Fig. 2.} {\it Distribution of a sample of 5,000
configurations generated by the MC simulation in the complex
plane of the magnetization for a lattice $T=R=18$. The
three clusters of points represent the one--phase configurations;
the straight lines joining the clusters are the two--interface
configurations.}
\vskip0.5cm
{\bf Fig. 3.} {\it The same of Fig. 2 for a lattice $T=10$, $R=20$.
The three-interface configurations are uniformly distributed
in the interior of the triangle.}
\vskip0.5cm
{\bf Fig. 4.} {\it Comparison of the predictions of the CWM with the
MC data: the energy gap $E$ is plotted as a function of
$z=R/T$. The lines represent the best fit of all
data to formula (\ref{e1}): from up to down they correspond
respectively to $T=9,10,11$ fixed (cfr. Tab. 1).}
\vskip0.5cm
{\bf Fig. 5.} {\it Comparison of the predictions of the classical
formula with the MC data: $E$ is plotted as a function of $A_m=RT$.
The line represent the best fit of the symmetric $T=R$ data, reported
with error bars, to formula (\ref{class});
the asymmetric MC data are also reported: squares correspond to
$T=9$ data, circles to $T=10$ and diamonds to $T=11$. }
\vskip0.5cm
{\bf Fig. 6.} {\it Comparison of the predictions of the two--loop
(\ref{e1}) and  one--loop (\ref{1lupo}) approximations for
the $T=11$ data, with $z=R/T$.
The dashed line represent the one--loop best fit while the two--loop
is the dotted line.}
\vskip0.3cm
\end{document}